\documentstyle[12pt,dina4]{article}
\newcommand{\ndelta}{$N + \Delta$ }
\newcommand{\deldel}{$\Delta + \Delta$ }
\newcommand{\sgommd}{$\sigma - \omega$ model~}
\newcommand{\NeNaF}{Ne~$+$~NaF~}
\newcommand{\NeNe}{Ne~$+$~Ne~}
\newcommand{\ArKCl}{Ar~$+$~KCl~}

\newcommand{\AuAu}{Au~$+$~Au~}
\newcommand{\Elab}{$E_{lab}$}
\newcommand{\PX}{$<p_X/A>$~}
\newcommand{\PY}{$<|p_Y|/A>$~}
\newcommand{\Kp}{$K^+$}

\newcommand{\eq}{\label}

\newcommand{\Pitild}{\mbox{${\tilde{\Pi}}$}}

\newcommand{\epsi}{\mbox{$\varepsilon$}}
\newcommand{\barg}{\mbox{$\bar{g}$}}

\newcommand{\tpsp}{\hspace{2.5em}}

\title{\Large \bf
Study of High Energy Heavy-Ion Collisions \\
in a Relativistic BUU-Approach  \\
with Momentum-Dependent Mean-Fields%
\footnote{Work supported by BMFT and GSI Darmstadt}}
\author{Tomoyuki Maruyama\thanks{Present Address: Department of Physics,
Kyoto University, Kyoto 606-01, Japan},
Wolfgang Cassing,\\
Ulrich Mosel, Stefan Teis and Klaus Weber  \\
\\
Institut f\"ur Theoretische Physik, Universit\"at Giessen, \\
D-35392 Giessen,  F. R. Germany }
\date{}

\begin{document}
\maketitle
\begin{abstract}
\noindent
We introduce momentum-dependent scalar and vector fields into the Lorentz
covariant relativistic BUU- (RBUU-) approach
 employing a polynomial ansatz for the
relativistic nucleon-nucleon interaction.
The momentum-dependent parametrizations
 are shown to be valid up to about 1 GeV/u
for the empirical proton-nucleus optical potential.
We perform  numerical simulations for heavy-ion collisions within
the RBUU-approach adopting  momentum-dependent and momentum-independent
mean-fields
and calculate the transverse flow in and perpendicular to the
reaction plane, the directivity distribution  as well as subthreshold
$K^+$- production.  By means of these
observables we discuss the particular role of the momentum-dependent
forces and their implications on the
nuclear equation of state. We find that only
a momentum-dependent parameter-set can explain the experimental data on
the transverse flow in the reaction plane from $150 - 1000$ MeV/u and the
differential \Kp- production cross sections at 1 GeV/u at the same time.
\end{abstract}

\section{Introduction}

\tpsp

The main aim of high energy heavy-ion physics is
to determine the equation of state (EOS) of nuclear matter
under extreme conditions far from the ground state.
To achieve this aim experiments between 100 $-$ 2100 MeV/u have been
performed  at the BEVALAC \cite{Naga} providing first information on
the properties of hot and dense nuclear matter.
A new generation of experiments is presently being performed at the SIS
\cite{SIS} in Darmstadt delivering more precise and exclusive data
about collective flow, multifragmentation or subthreshold particle production.

Any conclusion on the properties of hot and dense matter must rely on
the comparison of the experimental data with  theoretical predictions based on
nonequilibrium models. Among these,
the BUU-approach \cite{Bert,BUU1} is a very successful
approach in describing the time-dependent evolution of the complex system
especially with respect to single-particle observables.
As a genuine feature of transport theories it
has two important ingredients: the mean-fields or self-energies for nucleons
and an in-medium nucleon-nucleon cross-section that accounts for the
elastic and inelastic channels.
By varying the mean-fields $-$ which reflect a certain EOS $-$ and
comparing the theoretical calculations with the experimental data,
one expects to be able to determine the nuclear EOS \cite{BUU3}.

Within the framework of BUU-simulations we have succeeded to predict/
reproduce particle production data in heavy-ion collisions and to clarify
their reaction processes \cite{BUU4,Mosel}.
In spite of this success it has not yet been possible to determine the nuclear
EOS because
the mean fields are not uniquely fixed by the equation of state
alone. In addition,  it is so far not possible to extract
the nuclear EOS from the results
of the BUU calculations  without ambiguities for other model inputs.

The most important model input besides the nuclear incompressibility is the
momentum-dependence of the mean-fields; the nucleon-nucleus potential at normal
density is
attractive below and repulsive  above about 200 MeV, respectively.
Gale et al. \cite{Gale1} were the first to take the momentum-dependence of the
mean-fields into account in the BUU approach, but have
 not found any significant
effect on the transverse flow in the reaction plane.
Aichelin et al. \cite{Aich1}, using the Quantum Molecular Dynamics
 (QMD) model,
have found that the momentum-dependence plays a dominant
role e.g. for $K^+$ production \cite{Aich2}.
In addition, Li et al. \cite{Momd} also have obtained similar results in
particle production calculations for light nuclei within the QMD approach.

However, there are problems with momentum-dependent (MD) mean-fields
in these treatments that stem from  their
behavior above several hundred MeV/u where relativistic effects should
become significant.
Above 1 GeV/u, e.g., the Lorentz contraction of the phase-space distribution
as well as relativistic kinematics largely influence the observables of
interest.
The Lorentz covariance of the interaction also influences the results such as
the transverse flow \cite{MARU2} because the interaction range is Lorentz
contracted, too.
In addition, the Lorentz covariant vector-coupling produces a Lorentz force
directed sideways from the vector fields which is not accounted for in present
nonrelativistic transport simulations.
Thus the Lorentz covariance of the interaction also generates a qualitatively
new (i.e. directed) momentum-dependence of the mean-fields in the fast moving
system.
In the energy regime considered here we thus have to employ a Lorentz
covariant transport approach with realistic MD mean-fields to examine the role
of the MD forces as well as to determine the nuclear EOS.

The relativistic \sgommd \cite{Serot} and the Dirac phenomenology \cite{Hama1},
which are based on the same picture, have successfully explained various
properties of ground state nuclei \cite{Serot} as well as high energy
proton-nucleus scattering \cite{Hama1,Hama2}, and thus provide a realistic
basis for our investigation.
The analysis of the experimental data clearly shows that we have to deal
with large attractive scalar and repulsive vector fields in the medium.
The relativistic BUU- (RBUU-) approach \cite{BUU1,RBUU0,RBUU1} is constructed
in a Lorentz covariant way by combining the BUU-approach with the \sgommd
along the line of time-dependent Dirac-Brueckner theory  \cite{BUU1,DBHF1}.

Even without any explicit momentum-dependence of the scalar and vector
mean fields a momentum-dependence of the Schr\"odinger equivalent potential
automatically emerges as a consequence of the Lorentz transformation properties
of the vector-fields (see discussion below).
Bl\"attel et al. \cite{RBUU1,Blaet} have shown in the RBUU approach,
that this momentum-dependence (or the related effective mass) has a dominant
effect on the transverse flow in the reaction plane through the Lorentz force
from the vector fields.

Because of these large effects of a proper description of relativity on
the flow observables it is interesting to look also for possible effects
on particle production.
In fact, Lang et al. \cite{LangK} have shown in the RBUU approach that
the nuclear incompressibility $K$ is more important than the effective mass
for subthreshold \Kp- production in \AuAu collisions whereas
the conclusion is opposite in light systems such as \NeNe.

In all of these calculations, however, the strength of the
momentum-de\-pen\-den\-ce of the Schroedinger-equivalent optical potential
has been overestimated
at high relative energies because these calculations were based on the
original $\sigma - \omega$ model which contains no explicit momentum-dependence
of the nucleon self-energies \cite{BUU1,RBUU1}. This shortcoming is
essential because,
as discussed above, the transverse momentum distribution in heavy-ion
collisions has been found to be dominated by the MD potentials and to be
less sensitive to the incompressibility $K$ \cite{RBUU1,Blaet}.
Thus a proper treatment of these fields is a necessary prerequisite on the way
to determine the EOS of nuclear matter.

In preceding publications \cite{KLW1,KLW2} we have addressed the
latter problem more generally and proposed a Fock-term like
ansatz as well as suitable parametrizations  for the
MD interaction in line with the empirical optical potential
or results from Dirac-Brueckner calculations.
Sehn et al. \cite{Sehn} have attempted another approach for the MD mean-fields.
They have introduced  configuration-dependent couplings and fitted them to
DBHF results for thermalized nuclear matter as well as
two-Fermi-sphere configurations.
However, this latter approach is strictly applicable only to these limiting
configurations whereas our concept can be used  for
arbitrary phase-space distributions.

This explicit momentum-dependence is not very effective for nuclear matter
problems at zero temperature.
In matter-on-matter collisions at relativistic energies, however, it dominantly
affects properties of this system such as the binding energy,
the longitudinal and the transverse pressure \cite{KLW2}.
Especially the transverse pressure is found to be reduced at high bombarding
energies with respect to momentum-independent (MID) fields.
Thus the momentum-dependence plays an important role if the local momentum
distribution differs significantly from a spherical,
equilibrated configuration.

In a previous letter \cite{TOMO1} we have presented the first time-dependent
covariant simulations with MD scalar and vector fields and have studied
especially the transverse flow in nucleus-nucleus collisions at bombarding
energies from $E_{LAB} = 150 - 800$ MeV/u.
Our new approach has been found to reproduce the experimental flow data
\cite{EXP1,EXP0} quite successfully.
In that work, however, we have not obtained any definite information about
the nuclear EOS.

In this paper, therefore, we calculate further independent  observables
with the same model input and compare them with experimental data.
The most promising candidate is at present the subthreshold $K^+$- production
in heavy nucleus-nucleus collisions.
In fact, the previous RBUU \cite{LangK} and QMD calculations \cite{Huang}
have predicted a high sensitivity to the incompressibility  in \AuAu
collisions at $E_{lab} = 1$ GeV/u; however, reliable MD mean-fields have not
been employed so far.
In order to reduce the ambiguity of the input mean-field parametrizations,
we also calculate flow phenomena, in particular the directivity distribution
as suggested by Alard et al. \cite{Alard,Cass92},
using the same theoretical input for all observables.

The present paper is arranged as follows.
In Sec. 2 we briefly explain the RBUU approach and the methods
to introduce the MD scalar and vector mean-fields into this approach.
In Sec. 3 we calculate various observables: the transverse flow in the reaction
plane and perpendicular to this plane, the directivity distribution as
well as differential \Kp-production cross sections for \NeNaF and \AuAu.
The transverse flow and the subthreshold \Kp- yields are compared
with the most recent experimental data from the SIS \cite{EXPK}.
In order to minimize the ambiguities we aim at a set of mean-fields which
reproduce the experimental data on the transverse flow and the \Kp- production
at the same time.
We summarize and conclude our work in Sec. 4.

\section{Relativistic BUU Approach with Momentum Dependent Mean-Fields}

\tpsp

In this section we give a brief description of the RBUU-approach with the
MD mean-fields and present the actual simulation method
used for heavy-ion collisions.

In Ref. \cite{KLW1} we have formulated the Relativistic BUU equation with
MD mean-fields as
\begin{equation}
\{ [ \Pi_\mu - \Pi_\nu ( {\partial}^{p}_{\mu} U^{\nu} )
- M^* ( {\partial}^{p}_{\mu} U_s ) ]  \partial^\mu_x  -
[ \Pi_\nu ( {\partial}^{x}_{\mu} U^{\nu} ) +
M^* ( {\partial}^{x}_{\mu} U_s ) ] \partial^\mu_p \} f(x,p) = I_{coll} ,
\eq{e000}
\end{equation}
where $f(x,p)$ is the Lorentz covariant phase-space distribution function,
$I_{coll}$ is a collision term given in Ref. \cite{BUU1}, and $U_s$ and
$U_\mu$ are MD scalar and vector mean-fields, respectively.
Then the kinetic momentum $\Pi_\mu$ and effective mass $M^*$ become local
quantities and are defined in terms of the fields by
\begin{eqnarray}
\Pi_\mu (x,p) & = & p_\mu - U_\mu (x,p)
\\
M^* (x,p) & = & M + U_s (x,p) ,
\eq{e200}
\end{eqnarray}
which satisfy the mass-shell condition:

\begin{equation}
V(x,p) f(x,p) = 0
\eq{e210}
\end{equation}
with
\begin{equation}
V(x,p) \equiv \frac{1}{2} ( \Pi^2 (x,p) - M^{*2} (x,p) ) ,
\eq{e220}
\end{equation}
where $M$ and $p_\mu$ are the baryon mass and the canonical momentum,
respectively.
The above mass-shell condition (\ref{e210}) shows that $f(x,p)$ is
nonvanishing only for  $V = 0$.
The usual Wigner function ${\tilde f}(t;{\bf x},{\bf p})$, which is a
Lorentz scalar, is then obtained as
\begin{equation}
f(x,p) = {\tilde f}(t;{\bf x},{\bf p}) \ 2 {\Theta}(\Pi_0) \
\delta( \Pi^2 - M^{*2} ) ,
\eq{e230}
\end{equation}
where the step function $\Theta (\Pi_0)$ restricts the phase space to the
positive energy states.

In actual simulations the RBUU eq. (\ref{e000}) is solved with the
test-particle method \cite{TESTP} as follows:
the Wigner function ${\tilde f}(t;{\bf x},{\bf p})$ is expanded as
\begin{equation}
{\tilde f} (t;{\bf x},{\bf p}) = \frac{(2 \pi)^3}{ 4 {\tilde N}_T }
\sum_{i=1}^{ {\tilde N}_T \cdot A} \delta( {\bf x} - {\bf x}_{i}(t) )
\delta( {\bf p} - {\bf p}_{i}(t) ) ,
\eq{e240}
\end{equation}
with the total nucleon number $A$ and the number of test-particles per
nucleon ${\tilde N}_T$; the limit  ${\tilde N}_T \rightarrow \infty$
then provides the exact solutions.

Substituting the above eq. (\ref{e240}) into the RBUU eq. (\ref{e000}),
we can obtain the equations of motion for the test-particles as
\begin{eqnarray}
\frac{d {\bf x}_i}{d t} & = & {\bf D}_{p_i}
\epsi ({\bf x}_i, {\bf p}_i )
= \frac{ {\tilde{\bf \Pi}}_i }{ {\Pitild}_{0 i} }
\nonumber \\
\frac{d {\bf p}_i}{d t} & = & - {\bf D}_{x_i}
\epsi ({\bf x}_i, {\bf p}_i ) .
\eq{e250}
\end{eqnarray}
where $\epsi ({\bf x}_i, {\bf p}_i)$ is the on-mass-shell energy of the i-th
test-particles, and $\Pitild_\mu (x,p)$ is defined as

\begin{equation}
{\Pitild}_{\mu} (x,p) \equiv \frac{\partial}{\partial p^\mu} V(x,p) .
\eq{e260}
\end{equation}
In the above equations the partial derivative $\partial$ is defined by using
$p_0$ as an independent variable whereas the total derivatives ${\bf D}$
are defined using $p_0 \equiv \epsi ({\bf x}, {\bf p})$.

Next we give the parametrization of the MD mean-fields
and explain our implementation in the RBUU-simulation.
Following Ref. \cite{KLW1} we first separate the mean-fields into a local part
and an explicit MD part, i.e.
\begin{eqnarray}
U_s(x,p) & = & U_s^H (x) + U_s^{MD} (x,p) ,
\nonumber \\
U_{\mu}(x,p) & = & U_{\mu}^H (x) + U_{\mu}^{MD} (x,p)  ,
\eq{e300}
\end{eqnarray}
where the local mean-fields are determined by the usual Hartree equation:
\begin{eqnarray}
U_s^{H} (x) & = & - g_s \sigma_H (x) ,
\nonumber \\
U_\mu^{H} (x) & = & g_v \omega^H_\mu (x)
\eq{e310}
\end{eqnarray}
with
\begin{eqnarray}
\frac{\partial}{\partial \sigma_H} {\tilde U} [ \sigma_H (x)]
= g_s \rho_s (x)
\nonumber \\
m_v^{2} \omega_\mu^H (x) = g_v j_\mu^H (x) ,
\eq{e320}
\end{eqnarray}
where
\begin{equation}
\tilde{U} [ \sigma_H ]
= \frac{ \frac{1}{2} m_s^2 \sigma_H^2 + \frac{1}{3} B_s \sigma_H^3
+ \frac{1}{4} C_s \sigma_H^4 }
{ 1 + \frac{1}{2} A_s \sigma_H^2 }  .
\eq{e321}
\end{equation}
In the above equations the scalar density $\rho_s (x)$ and the current
$j_\mu^{H} (x)$ are given in terms of the phase-space distribution
function by

\begin{eqnarray}
\rho_s (x) & = &
\frac{4}{(2 \pi)^3}
\int d^4 p {~} M^* (x,p) f(x,p)
= \frac{4}{(2 \pi)^3} \int d^{3} p {~}
\frac{M^* (x,p)}{\Pitild_0 (x,p)} {\tilde f}(x,{\bf p}) ,
\nonumber \\
j_\mu^{H} (x) & = & \frac{4}{(2 \pi)^3} \int d^4 p {~} \Pi_\mu (x,p) f(x,p),
= \frac{4}{(2 \pi)^3} \int d^{3} p {~}
\frac{\Pi_\mu (x,p)}{\Pitild_0 (x,p)} {\tilde f}(x,{\bf p}) .
\eq{e330}
\end{eqnarray}

In eq. (\ref{e321}) we have introduced a new parameter $A_s$ in the denominator
of the sigma-field potential $\tilde{U}(\sigma_H)$ to avoid numerical
instabilities of the solution.
In previous applications, not including the denominator part
(i.e. $A_s \equiv 0$),
the value of $C_s$ sometimes could become highly negative, especially for low
incompressibility and low effective mass.
In this case the solution of eq. (\ref{e320}) cannot be uniquely determined
and the numerical simulation becomes unstable.

Now we present the actual parametrizations of the MD parts in eq. (\ref{e300}).
In order to be able to formulate a conserving theory (with respect to
the energy-momentum tensor) these parts are constructed
in analogy to Fock terms of nucleon self-energies.
Instead of adopting an infinite sum of Fock-like terms as in Ref. \cite{KLW1},
we use a simple parametrization for the MD parts as
\begin{eqnarray}
{U_s}^{MD} (x,p) & = & - \frac{4}{(2 \pi)^3}
\frac{{\bar g}_s^2}{m_s^2} \int d^{4} q {~} f(x,q) D_s(p,q) ,
\nonumber \\
& = & - \frac{4}{(2 \pi)^3} \frac{{\bar g}_s^2}{m_s^2} \int d^{3} q {~}
\frac{M^* (x,q)}{\Pitild_0 (x,q)} {\tilde f}(x, {\bf q}) D_s(p,q) ,
\nonumber \\
{U_\mu}^{MD} (x,p) & = & \frac{4}{(2 \pi)^3}
\frac{{\bar g}_v^2}{m_v^2} \int d^{4} q {~} {\Pi_\mu} f(x,q) D_v(p,q)
\nonumber \\
& = & \frac{4}{(2 \pi)^3} \frac{ {\bar g}_s^2 }{ m_s^2 } \int d^{3} q {~}
\frac{\Pi_\mu (x,q)}{\Pitild_0 (x,q)} {\tilde f}(x, {\bf q}) D_s(p,q) ,
\eq{e340}
\end{eqnarray}
with
\begin{equation}
D_{s,v} (p,q) = \frac{ \Lambda^2_{s,v} }{ \Lambda^2_{s,v} - ( p - q )^2 },
\eq{e350}
\end{equation}
where the effective coupling constants ${\bar g}_{s,v}$ and the effective
meson masses $\Lambda_{s,v}$ are treated as free parameters which are
fixed approximately by the empirical optical potential.

For the effective approach above (\ref{e340}), (\ref{e350}) the MD
parametrizations require some comments.
First, the above parametrization gives a reliable momentum-dependence of the
mean fields for  general momentum distributions as well as for the special
case of a single Fermi distribution because the expression (\ref{e340}) is
general within the Hartree-Fock theorem.
Second, our approach can reproduce the Dirac Brueckner results for
the scalar and vector fields, too \cite{KLW1}.
Third, comparing our parametrizations of Fock diagrams with vacuum polarization
calculations \cite{Henn}, we find that we
do not have any serious problem with the imaginary parts of the
nucleon self-energies, except for
off-mass-shell particles with $p_0 \gg \epsi({\bf p})$ which, however,
do not appear in semi-classical transport approaches according to the condition
(\ref{e210}).
Therefore we can neglect a more complicated density dependence for the MD parts
as well as vacuum polarization effects.

However, the above parametrization is still difficult to be applied
in actual RBUU-simulations \cite{KLW2}.
In principle we should use a six-dimensional grid in phase space for
the fields to get exact results,
but this method needs a large amount of memory and consumes tremendous CPU
times.
In order to avoid this difficulty, we propose an alternative polynomial
approximation for the propagator $D_{s,v}$ as
\begin{eqnarray}
D_{s,v} (p,q) & = & 1 + \frac{a_{s,v}}{m_{s,v}^2} ( p - q )^2
                  + \frac{b_{s,v}}{m_{s,v}^4} ( p - q )^4 .
\eq{e360}
\end{eqnarray}
In this way we obtain analytical expressions for the MD fields
as
\begin{eqnarray}
U_s (x,p) & = & - g_s \sigma_H (x) - \frac{\barg_s^2}{m_s^2} \rho_s (x)
+ p^2 U_s^{(1)}(x) + p^4 U_s^{(2)}(x)  ,
\nonumber \\
U_{\mu} (x,p) & = & \frac{( g_v^2 + \barg_v^2 )}{m_v^2} j_\mu^{H} (x)
+ p^2 U_\mu^{(1)}(x) + p^4 U_\mu^{(2)}(x)  .
\eq{e370}
\end{eqnarray}
The spatial derivatives of $U_{s}$ and $U_{\mu}$ at fixed momentum are easily
evaluated with the above expressions.
The above polynomial approximation is also applicable for
general configurations below a certain limiting energy as will be discussed
below.
Note that eq. (\ref{e370}) does not reduce to MD coupling parameters as in
Ref. \cite{Sehn} since each term has a different $x-$dependence.
The conserving character of the theory \cite{KLW1} is thus maintained.

The free parameters are determined in the following way:
we fit the mean-fields using the ansatz (\ref{e350}) to reproduce the
experimentally observed nucleon-nucleus potential and then determine the
parameters of the polynomial approximation (\ref{e360}) to give the same
results
for the mean-fields up to about 1 $-$ 1.2 GeV  kinetic energy of the nucleon.
The resulting parameters are given in Table 1.
All parameter-sets yield a saturation density $\rho_0 = 0.17 {\rm fm}^{-3}$
and a binding energy per nucleon of $E_B / A = - 16 {\rm MeV}$ at $\rho_0$.
The incompressibility $K$ and the
effective mass $M^* / M$ at the Fermi level and $\rho = \rho_0$ are
also given in Table 1.

In order to compare our calculations with momentum-independent mean-fields,
furthermore, we also present the parameter-set NL6, NL7, NL21 in Table 1
where the corresponding nuclear matter properties are also given
in the bottom lines.
Note that the above saturation density is different from that for the
NL parameter-sets used in the previous RBUU-approach
\cite{RBUU0,RBUU1,Blaet,Koch,LangK}.
For the old NL parameter-sets the value of the saturation density had been
taken as $\rho_0 = 0.145 {\rm fm}^{-3}$, which is too small in comparison
to the bulk density of heavy nuclei.
The old NL parameter-sets lead to vector fields which are about 10\% larger
than the present ones; this difference influences especially calculations
at high density and/or high energy.
Thus we have refitted the parameter-sets to reproduce the proper saturation
density in all cases.

In Fig. 1 we show the resulting MD mean-fields $U_s$ (a) and $U_0$ (b) at
normal nuclear matter density $\rho = \rho_0$ as a function of the nucleon
kinetic energy $\epsi_K$.
Solid, thick dashed and dashed lines indicate the results of POL6, MD6 and NL6,
respectively.
By construction the two results of POL6 and MD6 are almost the same up to
1.2 GeV/u while the polynomial approximation fails at higher energy.
This limits the applicability of the polynomial approximation to energies
up to about 1 GeV/u.

In Fig. 1c we present the Schr\"odinger equivalent potential defined by
\begin{equation}
U_{SEP} = U_s + U_0 + \frac{1}{2 M} (U_s^2 - U_0^2 ) + \frac{U_0}{M}
\varepsilon_{K}.
\eq{e380}
\end{equation}

\noindent
The hatched area indicates the result of the experimental analysis by
Hama et al. \cite{Hama2}.
Again, the optical potentials for the parametersets MD6 and POL6 are almost
identical and approximately reproduce the experimental result.
The $U_{SEP}$ of NL6 is linear in $\epsi_K$ and not so
different from the experimental result up to about 300 MeV, but yields
too strong repulsion at higher energy (cf. Sec. 1).
Since this strong repulsion is due to very strong vector fields,
the NL6 parameter-set overestimates the Lorentz force on a fast moving
particle in the medium.

In Fig. 2 we show the energy per nucleon for nuclear matter (a) and
the effective mass at the Fermi surface (b) as a function of density
for each parameter-set.
The expression for the total energy has been given in Ref. \cite{KLW1}.
Again, the EOS of POL6 (solid line) almost completely agrees with that of
MD6 (thick dashed line). Besides, it is not very different from that
for NL6 (dashed line) since the explicit momentum-dependence does not
significantly change the nuclear EOS at moderate densities.
In Fig. 2, furthermore, we display the results for POL7 (dash-dotted line),
NL7 (dotted line) and NL21 (thick dotted line),
which are discussed in Sec. 3.
Here we only  confirm again that the momentum-dependence does not strongly
change the nuclear matter properties.

In Ref. \cite{KLW2}, however, it was shown that the momentum-dependent
forces have a large influence on matter-on-matter collisions where the momentum
distribution is described by two shifted Fermi spheres.
Before performing actual simulations, we compare our polynomial
approximation to the exact results for the above configuration.
The negative binding energy (a), longitudinal (b) and transverse pressure (c)
are shown for all parameter-sets in Fig. 3 as a function of the bombarding
energy per nucleon.

First, the results of POL6 (solid line) reproduce those of MD6
(thick dashed line)
up to the bombarding energy of about \Elab = 1 GeV/u for all three quantities;
the biggest difference shows up in the longitudinal pressure at the highest
energy ($\Delta p_L < 10 \%$).
Thus we can safely conclude that the polynomial approximation should be valid
up to this energy in both proton-nucleus and nucleus-nucleus collisions.
Second, the energy and the longitudinal pressure are almost completely
determined by the momentum-dependence.
Note especially the big difference in the binding energy between MD and
MID parametrizations.
Third, the negative binding energy is almost proportional to the bombarding
energy \Elab \ in the MID parameter-sets and saturates in the high-energy
region
above about \Elab = 300 $-$ 400 MeV/A for the realistic MD parametrizations.
Fourth, the MD parametrizations reduce the transverse pressure at high
bombarding energies  around 1 GeV/u quite significantly; on the other hand,
below about 500 MeV/u a sizeable dependence
on the incompressibility  can be seen.
Thus the dynamical effects of the MD mean-fields should show up in the early
stage of a heavy-ion collision when the momentum distribution is almost
equivalent to two shifted Fermi spheres.

The most drastic effects of the momentum-dependence arise for the negative
binding energy and the transverse pressure where the results obtained
with the correct momentum-dependence are much lower than those calculated with
the original \sgommd.
The lowering of the transverse pressure and the negative binding energy is a
direct consequence of the weakened strength of the Lorentz force at
higher energy.
{}From the above results we can already predict that the MD mean-fields
in actual heavy-ion collisions  will lead to a higher compression (as a
consequence of the smaller total energy) and a suppression of the transverse
flow (due to a smaller transverse pressure) as compared to the MID
parametrizations.

\section{Results and Discussion}

\tpsp

In this section we report results of
the actual numerical simulation of the RBUU approach
with the MD mean-fields and calculate various
observables for heavy-ion collisions.
We use 150 $-$ 1000 testparticles per nucleon in the simulation
and adopt the Cugnon parametrizations of the baryon-baryon cross-sections
\cite{Bert}, \cite{Cug} for the stochastic baryon-baryon (BB)
 collisions including
the elastic and inelastic channels with the $\Delta$ resonance in the
"frozen $\Delta$" approximation.
In the actual simulation the collision term is treated within the Local
Ensemble Method described in Ref. \cite{Local}, which gives a proper solution
of the BUU equation.

Before calculating observables we examine the time dependence of the maximum
density in the local rest frame ($\rho_r = \sqrt{j_\mu j^\mu}$) in \AuAu
collisions for the parameter-sets POL6, POL7, NL6, NL7 and NL21 at the
bombarding energy \Elab = 400 MeV/u for the impact parameter $b$ = 3 fm (a)
and at \Elab = 1 GeV/u for $b$ =  0 fm (b) (Fig. 4).
The parametrizations POL6 and NL21 lead to much higher densities than NL6 and
NL7 at \Elab = 1 GeV/u, whereas the  parameter-set NL7 produces
only a slightly higher
density than NL6 because the EOS for NL7 is softer than that for NL6
(cf. Fig. 2a).
These tendencies are also present at \Elab = 400 MeV/u, but cannot
be identified so clearly.
Since the MD mean-fields reduce the repulsion in the high energy region,
the mean force between two nuclei is much weaker for POL6 and POL7 than
for NL6 and NL7 at the beginning of the nucleus-nucleus collisions.
In addition, we can understand the similarity of the results obtained with POL6
and NL21 at \Elab = 1 GeV/u from the fact that the MD parameter-set POL6
generates the same strength of the Schr\"odinger equivalent potential as
NL21 around 1 GeV/u (cf. Fig. 1c).
Thus the maximum density in the compression phase is determined by
the property of the mean-fields at the early times of the reaction.

\newpage

\subsection{Collective Flow}

\tpsp

We have already discussed that the explicit momentum-dependence of the mean
fields decreases the transverse pressure in the system for the two-Fermi-sphere
geometry which is realized in the early stage of a heavy-ion collision.
As discussed in the previous section, this fact predicts also a
suppression of the transverse flow. To demonstrate this
we calculate the mean transverse momentum in the reaction plane \PX
versus the normalized center-of-mass (CM) rapidity $Y_{CM}/Y_{PR}$
($Y_{PR}$: the projectile rapidity),
which has been shown in Ref. \cite{RBUU1,Blaet} to be very sensitive to the
momentum-dependence, but almost insensitive to the incompressibility $K$.

Fig. 5 shows the results in \ArKCl collisions at \Elab = 800 MeV/u (a),
\AuAu collisions at \Elab = 400 MeV/u (b) and 150 MeV/u (c)
at b = 6 fm for the parametrizations POL6 (solid lines), POL7
(dash-dotted lines), NL6 (dashed lines) and NL7 (dotted lines).
For the calculation of \AuAu collisions at 400 MeV/u
the (PlasticBall) filter routine
and the coalescence model of \cite{RBUU1} are used.

The differences between our various parametrizations are negligible
at \Elab = 150 MeV/u in the interesting regime around midrapidity,
but become larger with increasing beam-energy.
At \Elab = 400 MeV/u, both the momentum-dependence and the incompressibility
lead to differences at high rapidity $Y_{CM}/Y_{PR} > 0.6$, however,
there are again no sizeable effects at midrapidity.
The parameter-sets  POL6 and NL7 give similar results whereas the flow angle
($\theta_{flow}$) for NL6 is bigger than for POL6 and that for POL7 smaller
than for NL7; i.e. $\theta_{flow}(POL7) < \theta_{flow}(NL7) \approx
\theta_{flow}(POL6) < \theta_{flow}(NL6)$.
Thus a low incompressibility and a low momentum-dependence clearly reduce the
transverse momentum, as expected; in addition, the behavior of the above
results
is quite similar to that of the transverse pressure for the two-Fermi-sphere
geometry around 400 MeV/u (cf. Fig. 3c).
However, these differences are not very pronounced because all results
almost agree with
the experimental data of the plastic ball group \cite{EXP0}.

At higher energy (800 MeV/u) the MD mean-fields  POL6, POL7 nicely
reproduce the experimental data \cite{EXP1} (a) while the flow for the
MID parameter-sets NL6 and NL7 is overestimated at 800 MeV/u;
the results are quite insensitive to the incompressibility $K$.
This behavior is again found to be similar to the transverse pressure
in the two-Fermi-sphere geometry around 800 MeV/u.

{}From the above results we learn that the  transverse flow is strongly
correlated with the transverse pressure for the two-Fermi-sphere geometry
as it exists in the early stages of the collision.
Both the momentum-dependence and the incompressibility contribute to the flow
around \Elab = 400 MeV/u. At the highest energy, however,
the momentum-dependence  dominates.
Note that this  holds in spite of the fact that
the densities reached here are even higher than at the lower energies.
This demonstrates again (cf. Ref. \cite{Blaet}) that it is not the density
compression but the Lorentz force which drives the transverse flow.
At \Elab = 800 MeV/u the MD parametrization POL6 and POL7 can reproduce the
experimental results whereas the results of the MID parametrization NL6 and NL7
overestimate the data.
The latter result is due to the fact that our MD mean-fields (POL6 and POL7)
yield the proper strength of the Lorentz force while the mean-field of NL6
and NL7 yield a too strong repulsion at high energy (cf. Fig. 1c).

We can thus not obtain any precise information about the nuclear EOS
from \PX, which is too insensitive to the incompressibility
in the high energy region.
The reason for this insensitivity is that the quantity \PX is
essentially produced by the Lorentz force \cite{Blaet} in the surface
region of the overlapping nuclei in the initial phase,
where the density is not so high and the momentum-distribution is
similar to the two-Fermi sphere geometry .
When increasing the bombarding energy, the distance between these two Fermi
spheres becomes larger and the properties of the system
are dominantly determined by the momentum-dependence of the mean-fields.
In order to determine the nuclear EOS, therefore, we need to discuss
also other observables from a heavy-ion collision.

Next we discuss the directivity distribution, which was suggested by
Alard et al. \cite{Alard} as a probe for the nuclear EOS.
The directivity $D$ is defined by

\begin{eqnarray}
D = \frac{ | \sum_i {\bf p}_T (i) | }{ \sum_i | {\bf p}_T (i) |}
& {\rm for} & Y_{CM} (i) > 0 ~~ {\rm and}~~
7^\circ < \theta_{lab} <  30^\circ.
\eq{e420}
\end{eqnarray}

\noindent
In Fig. 6 we show the yield distribution for the observable $D$ in a \AuAu
collision at 400 MeV/u for the impact parameter average $b \leq 5$fm.
We find differences with respect to the incompressibility as well as
to the momentum-dependence.
A low incompressibility or a low momentum-dependence shift the results
to smaller directivity, as expected.
The results for POL6 (hard EOS) and NL7 (soft EOS) agree with each other
though the maximum yield is slightly different.

In order to examine the directivity in more detail we show the average value
$<D>$ and its width ${\Delta}D$ in Fig. 7 as a function of impact parameter.
With increasing impact parameter the average value increases first, has a
peak at about $b = 5$ fm, and then decreases
again for very peripheral reactions.

It can be seen again that a weak momentum-dependence and a soft EOS reduce
both the average value and the width.
For larger impact parameters around $b = 7$ fm (comparing POL6 and POL7),
moreover, the effect of the incompressibility becomes almost negligible and
the momentum-dependence is dominant because there is no compressed  participant
zone in peripheral collisions \cite{Koch}.
Thus the directivity distribution might be useful to determine the nuclear EOS,
but only after MD mean fields have been determined from a study of
peripheral reactions.

In recalling: the directivity distribution is influenced both by the
incompressibility and the momentum-dependence.
The directed flow in the $x$- direction (cf. \PX) is still essential to
build up the directivity; consequently  there are
no clear differences in the small impact parameter region.
In addition, the behavior is also quite similar to that of \PX for large
rapidities.
Hence the directivity is again determined by the transverse pressure in the
early stage of the reaction, but the differences between various
input quantities, that show up only weakly in the mean transverse
 momentum \PX, are magnified.

We  leave the discussion on the directivity with a short comment:
the authors of refs. \cite{Alard,Coff} have
 suggested to select events from central
collisions using the restriction  $D < 0.2$ together with
a high  multiplicity bin.
Even taking into account the width in the directivity distribution, this
restriction should select central events with $b < 2$ fm in all cases;
our results thus strongly support their idea.

Next we discuss the mean absolute momentum perpendicular to the reaction plane
\PY defined as
\begin{equation}
< | p_{Y} | / A > = \frac{1}{{\tilde N}_T A} \sum_i | p_y (i) | .
\eq{e410}
\end{equation}
where the $y-$direction is defined as that perpendicular  to the
reaction plane ($x$, $z$).
In Fig. 8 we show the $<|p_{Y}| / A >$ versus the rapidity (a)
in \AuAu collisions
at \Elab = 400 MeV/u for central collisions ( $b < 3$ fm).
For reference we add the rapidity distribution in Fig. 8b.
The results for the parameter-sets POL6 and NL6 ($K$ = 400 MeV) are larger
than those for POL7 and NL7 ($K$ = 200 MeV) for small rapidities
$Y_{CM}/Y_{PR} < 0.3$, while there is no significant difference among all cases
in the rapidity distribution.

We thus find that the flow in y-direction is more sensitive to the
incompressibility than to the momentum-dependence.
This result can be understood as follows: On one hand, the spectators do not
contribute significantly to the flow in y-direction since the Lorentz force -
which is the dominant force for the spectators due to their high relative
velocity - acts in the reaction plane $(x, z)$. On the other hand,
the nucleons in the participant zone are accelerated by the pressure in
the high density participant region into all directions. These nucleons
experience many baryon-baryon (BB) collisions and consequently populate
the CM midrapidity
region which involves only relatively small momenta. Thus, for these nucleons
the Lorentz force is rather weak and their final momenta are mainly
determined by the incompressibility.
Unfortunately, the sensitivity of the flow
perpendicular to the reaction plane with respect to the
incompressibility is rather  weak;
the differences caused by the incompressibilty $K$ are about $5 - 10$ \% of the
total strength.

In summary, the results on \PX show that our MD parametrizations describe the
Lorentz forces in the initial stages of the collision correctly,
but they also show that the in-plane transverse flow does not give any
information on the nuclear EOS. This is in line with the results of
Lang et al. \cite{Lang} who showed that the in-plane transverse flow
is determined by the
transverse pressure, but does not depend sensitively on the EOS.
On the other hand, the directivity distribution
and the \PY can be used as probes to
determine the incompressibility and help to understand the global evolution
of the heavy-ion collision.
However, also these observables are not very sensitive to the properties
of highly compressed  matter.
In the high energy region (\Elab $~>$ 1 GeV/u) the transverse pressure of the
initial overlap area is almost entirely determined
by the momentum-dependence of
the mean-fields at $\rho = \rho_0$ and the contribution from the collision
terms becomes large, too.
Thus we have to study other observables such as particle production yields at
subthreshold energies as a trigger on high baryon density.

\subsection{Kaon Production}

\tpsp

In this section we now turn to the study of particle production at
subthreshold energies where the particles are predominantly produced in the
high density zone \cite{LangK}. We study these reactions with the hope to
narrow down the ambiguities in the nuclear EOS as obtained from the study
of flow phenomena. In particular, we
discuss in this section the \Kp- production in the subthreshold
energy region, which has been suggested by Aichelin and Ko to be a good probe
for the nuclear EOS \cite{AiKo}.
There are no serious ambiguities in the $K^+$  production process:
neither $N +$ multi$\times{N}$ \cite{Batko1} nor  $\pi + N$ collisions
\cite{Batko2} play a significant role; the dominant contribution comes from
binary BB collisions.
In addition, \Kp- production can be calculated perturbatively because of the
small
production probability in BB  collisions and the negligible absorption in
the medium due to strangeness conservation.
In Ref. \cite{LangK}  Lang et al. have found that this observable
should be  a good probe in \AuAu collisions at \Elab = 1 GeV/u.

Since the detailed description of the production calculations is given in
\cite{LangK}, we directly show the results of our present calculations for
the experimental angle $\theta_{lab} = 44^\circ$
in \NeNaF (a) and \AuAu (b) collisions at $E_{lab} = 1$ GeV/u (Fig. 9).
Also displayed in Fig. 9 are the experimental data from \AuAu collisions and
the preliminary data from \NeNaF collisions of the KAOS
 group at SIS \cite{EXPK}.
In the \NeNaF reaction there is no significant difference among all cases
except for the results for NL21 which are about 15 \% larger than the others.
The reason for this behavior has already been shown in Ref. \cite{LangK}:
in \NeNaF collisions only moderate densities in the medium ($\rho < 2\rho_0$)
are achieved such that the results do not reflect the properties of
high density nuclear matter.
Furthermore,  the \NeNaF results quite strongly depend on the surface
properties.
For example, when varying the surface thickness by about 25 \%,
the inclusive $K^+$ yields change by about a factor of 5 in the RBUU
simulation.
This fact shows that the total cross-section is determined by the initial
phase-space distribution and thus very sensitive to the initialization.
The strong sensitivity to the initialization is a typical property of
the deep subthreshold particle production in light systems where the
surface plays a larger role.
It does not occur for heavy systems; the different initializations change
the results for \AuAu only by a few percent.

{}From the \AuAu collisions we get some information about the high density
matter.
The calculated yield is sensitive both to the
incompressibilty and the effective mass
at the Fermi surface. However, the explicit momentum-dependence
does not change the spectrum very much
except for the slope which is slightly smaller for POL6 than for NL6.
This shows that the MD mean-fields generate
higher kinetic energies or higher temperatures than the MID fields.
The yield is ordered as $(NL21) > (NL7) > (NL6) \approx (POL6)$.
This order is equivalent to that in the binding
energy of high density nuclear matter.
The results for the $K^+$ yield thus reflect the energy density of
nuclear matter.

The MD parametrization POL6 can reproduce the experimental data for \Kp-
production as well as the transverse flow whereas none of the  MID
parametrizations can explain both sets of data at the same time.
Hence we expect that the nuclear EOS corresponding to the parameter-set
POL6 is very close to the exact one for $\rho \approx$ 2 $-$ 3 $\rho_0$.

Next we examine the contributions from the various production channels.
In Fig. 10 we show the results for NL6 from the $N + N$, \ndelta and \deldel
collisions separately.
We find that 90 \% of the kaons are produced from the \ndelta or
\deldel channels while the $N - N$ channel contributes only
10 \% to the total yield (cf. Ref. \cite{LangK}).
Furthermore, the slopes of each partial cross-section do not show any
remarkable differences; apparently the nucleon and the $\Delta$ feel a similar
temperature inside the compression zone.

In order to understand these results in more detail,
we investigate the dependence of the $K^+$ yield on the density
in the local rest frame and on the
baryon effective mass  for \NeNaF and  \AuAu collisions;
the results are shown in Fig. 11 as a function of the ratio
$r = {\rho_r}/\rho_0$ and $m = M^*/M$.
A first observation is that the shape of the density dependence $-$ not the
absolute magnitude $-$ is dominantly  determined by the momentum-dependence
and the effective mass at the Fermi surface; the NL6 and NL7 mean-fields yield
similar results and also POL6 and POL7.
Second, the results with the POL6 mean-fields show tails up to the high
density region ($\rho > 3\rho_0$);
in the \AuAu collision the maximum density is about 4$\rho_0$ for POL6 and
almost the same as that for NL21 (cf. Fig. 4).
As shown in Fig. 1, the momentum-dependence reduces the repulsion in the
initial stage and this permits higher densities than the MID mean-fields.
However, the peak position for POL6 agrees with those for
NL6 and NL7, while the peak position for NL21 is higher than the others.
The reason for these dependences is as follows:
The EOS of POL6 is almost equivalent to that of NL6.
The parameter-set POL6 leads to a higher density than NL6, but
nevertheless there is not enough energy to produce more
\Kp-mesons because the bombarding energy is consumed by the
build-up of the vector field;
consequently, \Kp-mesons cannot be easily produced in the compression zone.

Fig. 11b shows the distribution of the effective masses of the baryons
involved in the $K^+$ production. All curves exhibit two maxima, one at
a value significantly below $\rho_0$ and the other one approximately 0.3
GeV above. While the lower one reflects events from nucleon-nucleon collisions
taking place at high densities, the higher one stems
from the $\Delta$ resonances in a $\Delta-N$
collision event, respectively.  The curves
shown here do not correlate with the magnitude of the cross sections; there
is thus no obvious connection between the number of kaons produced and
the momentum-dependence of the interaction.

We briefly summarize the effects of the MD mean fields for \Kp- production.
The kaon yield is determined both by the collision rate and by the
momentum distribution in the participant zone. The former is directly
related to the maximum density reached and this, in turn, depends on the
behaviour of the mean fields at the initial high energy. The latter, on
the other hand, is determined by the low-energy behaviour of the mean
fields which becomes relevant after many BB collisions have taken
place. The MD parametrizations are much less repulsive than the MID ones
in the early stage of the collision and thus lead to higher densities, but
the incoming energy has partly been used to build up the
correspondingly higher vector field. Both effects counteract each other
to a large extent such that the kaon yield becomes rather insensitive
to the momentum-dependence. Thus $K^+$ production around 1 GeV/u is a
good tool to study the EOS.

Comparing with the results of ref. \cite{LangK} we see that the kaon
production cross sections are about a factor of two larger than those
obtained there (for NL21 versus NL2). This difference is due to the fact
that a lower saturation density was used in \cite{LangK}; this affects
the absolute yield but leaves the slope of the spectra nearly unchanged.
Finally we would like to note that our conclusion of a small effect of
the momentum dependence on the $K^+$ yield might seem to be at variance
with the results of QMD calculations in refs.  \cite{Aich1,Momd}, where
a large effect was obtained. However, the results of ref. \cite{Aich1}
show large statistical errors and have been obtained in an approach that
lacks manifest covariance in the mean fields thus limiting their
reliability in the relativistic energy domain treated here. The results of
ref. \cite{Momd} on the other hand are not inconsistent with ours; they
were performed only for light systems where the $K^+$ yields depend strongly
on the surface properties and on the initialization.

\section{Summary}

\tpsp

In this paper we have introduced  momentum-dependent (MD) mean-fields into
the RBUU approach in order to correct for the too repulsive potentials
at high energies in the original approach.
We have performed  actual simulations of heavy-ion collisions from
$E_{lab} = 150$ MeV/u to $1000$ MeV/u employing two momentum-dependent
parametrizations (POL6 and POL7) with a polynomial approximation for the
nucleon-nucleon interaction and three momentum-independent
parametrizations (NL6, NL7 and NL21).
The MD parameter-sets (POL6, POL7) are fitted to the empirical optical
potential for proton-nucleus scattering and lead to less repulsive
mean fields in the high energy region than the MID parameter-sets
NL6 and NL7.
As a consequence, also in nucleus-nucleus collisions the explicit
momentum-dependence
suppresses the repulsion in the high energy region
$E_{lab} > 300$ MeV/u; this effect is most pronounced in the early
stage of the collision process.  Due to the reduced
repulsion from the MD mean-fields the system achieves
higher densities  and a lower transverse pressure in comparison
to the results for the MID mean fields.
The effect becomes even more important  with increasing  bombarding energy.

We have examined a variety of observables in these RBUU simulations:
the mean transverse momentum in the reaction plane \PX, the directivity
distribution, the mean absolute transverse momentum perpendicular to the
reaction plane \PY and the subthreshold \Kp- production.
The quantity \PX is found to be sensitive to the
momentum-dependence, but much less
sensitive to the nuclear incompressibility.
The quantity \PY, on the contrary, is found to be
rather insensitive to the momentum-dependence. Finally, the
directivity distribution is both sensitive to the incompressibility and the
momentum-dependence; however, the differences between the results obtained
with the incompressibilities $K$ = 200 MeV and
$K$ = 400 MeV are too small to be observed experimentally.
On the other hand, the subthreshold \Kp- production in \AuAu collisions turns
out to be a suitable probe for the nuclear EOS, thus confirming the
original suggestion of ref. \cite{AiKo}. The differential yield is
sensitive to the nuclear incompressibility but insensitive to the
momentum-dependence for heavy systems like \AuAu.

{}From the comparison of the above results, furthermore, we obtain a clear
picture of the effects of the MD mean-fields.
The high energy behaviour of the mean-fields is significant in the initial
stage and the strength at the initial energy determines the mean
transverse flow and the density distribution in the participant zone.
With increasing time, the matter is compressed and the bombarding energy is
distributed among the nucleons in the high-density participant zone
where the average relative
momentum between two baryons becomes much smaller.
In the most condensed phase the high energy behaviour of the mean-fields
becomes less important and the low energy behaviour of the mean-fields
essentially  determines the energy-momentum distribution in the participant
zone due to energy-momentum conservation.
Thus the phase-space distribution in the participant zone is determined
by both: the high energy behaviour at $\rho = \rho_0$ and the low energy
behaviour of the mean-fields in the whole density region.
Fortunately, the difference in the maximum densities between POL6 and NL6
does not influence the result for \Kp- production.
In addition, only the nuclear incompressibility $K$  affects the $K^+$
yield significantly; the scalar and vector fields have no
separate effect on this quantity.

{}From the results of the present investigations and the comparison to various
experimental data on flow phenomena and kaon production simultaneously
we infer that the true nuclear EOS should be
close to that of POL6 around $\rho \approx 2 - 3 \rho_0$.
Essential for this conclusion is the simultaneous analysis of unrelated
sets of data within the same model
since there might be unknown medium effects as e.g.
suggested by Ko et al. \cite{Ko2} for the $K^+$ mesons and Jung et al.
\cite{Jung} for
long-range correlations induced by the vacuum polarization.


\newpage

\begin{center}
{\bf Table 1 }
\vspace{1em}

\begin{tabular}{|c|c|c|c|c|c|} \hline
                  &  POL6   &   NL6  &  POL7  &  NL7   & NL21    \\ \hline
 $g_s$            &  6.95   &  9.39  &  8.60  &  9.94  & 7.42   \\ \hline
 $g_v$            &  0.0    &  11.0  &  0.0   &  11.0  & 6.80  \\ \hline
 $B_s$ (fm)       & $-2.29$ &  2.95  &  43.7  &  19.8  & 0.0    \\ \hline
 $C_s$ (fm$^2$)   &  98.3   &  7.82  &  0.0   &  0.0   & 912    \\ \hline
 $A_s$ (fm$^2$)   &   0.0   &  0.0   &  13.6  &  5.75  & 59.4    \\ \hline
 ${\bar g}_s$     &  6.42   &        & 5.51   &        &         \\ \hline
 ${\bar g}_v$     &  11.3   &        & 11.3   &        &         \\ \hline
 $a_s$            & 0.276   &        & 0.267  &        &         \\ \hline
 $b_s$            & 0.0125  &        & 0.0150 &        &          \\ \hline
 $a_v$            &  0.276  &        & 0.267  &        &         \\ \hline
 $b_v$            & 0.0328  &        & 0.0306 &        &         \\ \hline
\hline
 $K $ (MeV)       &   400   &   400  &  200   &  200   &  200    \\ \hline
 $M^*/M$          &  0.65   &  0.65  &  0.65  &  0.65  &  0.83   \\ \hline
\end{tabular}

\vspace{2em}

\end{center}
\vspace{1em}

\noindent
Parameters for the various equations of state discussed in this paper.
Also given are the incompressibility $K$ and the effective mass $M*$.
In all cases we have used $m_s$ = 550 MeV and $m_v$ = 783 MeV.

\hfill

\eject

\newpage


\noindent
{\large\bf Figure captions}\hfill

\begin{description}
\item[Fig. 1]
The kinetic energy dependence of the scalar fields (a),
time-component of vector fields (b) and the Schr\"odinger equivalent
potential (c).
The results for POL6 and MD6 are denoted by solid and thick dashed lines,
respectively.
The dashed lines show the results for the momentum-independent mean-field
(NL6) and the thick dotted lines those for NL21.

\item[Fig. 2]
The equation of state for nuclear matter: total energy per nucleon versus
baryon density (a) and the nucleon effective mass divided by the
 vacuum mass (b).
The solid, dash-dotted, dashed, dotted , thick dotted and thick dashed lines
show the results for POL6, POL7, NL6, NL7, NL21 and MD6, respectively.

\item[Fig. 3]
The negative binding energy per nucleon (a), the longitudinal pressure (b) and
the transverse pressure (c) versus the kinetic energy
per nucleon in the Lab. frame. for matter-on-matter configurations (see text).
The solid, dash-dotted, dashed, dotted , thick dotted and thick dashed lines
show the results for POL6, POL7, NL6, NL7, NL21 and MD6, respectively.

\item[Fig. 4]
The time dependence of the maximum density (in units of $\rho_0$)
for \AuAu collisions at $E_{LAB}$ = 1GeV MeV/u and b = 0 fm (a),
and \AuAu at 400 MeV/u and b = 3 fm (b).
The solid, dash-dotted, dashed and dotted lines
show the results for POL6, POL7, NL6 and NL7, respectively.

\item[Fig. 5]
The mean transverse momentum per nucleon $p_x$/A versus
the center-of-mass rapidity (per initial projectile rapidity) for \ArKCl
collisions at $E_{LAB}$ = 800 MeV/u (a), \AuAu at 400 MeV/u (b) and
\AuAu at 150 MeV/u (c) at b = 6 fm.
The solid, dash-dotted, dashed and dotted lines
show the results for POL6, POL7, NL6 and NL7, respectively.
The experimental data (asterisks) are taken from \cite{EXP1} for (a) and
\cite{EXP0} for (b).

\item[Fig. 6]
The distribution of the directivity D in \AuAu collisions at 400 MeV/u for
$b < 5$ fm and the experimental acceptance $7^\circ <
\theta_{lab} < 30^\circ$.
The solid, dash-dotted, dashed and dotted lines show the results for POL6,
POL7, NL6, and NL7, respectively.

\item[Fig. 7]
The impact parameter dependence of the average directivity distribution (a)
and the width of the directivity distribution (b).
The solid, dash-dotted, dashed and dotted lines
show the results for POL6, POL7, NL6 and NL7, respectively.

\item[Fig. 8]
The mean absolute transverse momentum perpendicular to the reaction plane
versus the center-of-mass rapidity.
The solid, dash-dotted, dashed and dotted lines
show the results for POL6, POL7, NL6 and NL7, respectively.

\item[Fig. 9]
The inclusive cross-section for \Kp- production for \NeNaF (a) and \AuAu
collisions (b) at $\theta_{Lab} = 44^\circ$ and \Elab = 1 GeV/u.
The solid, dashed, dotted and thick dotted lines show the results for POL6,
NL6, NL7 and NL21, respectively.
The asterisks show the experimental data \cite{EXPK}.

\item[Fig. 10]
The inclusive cross-section for \Kp- production for \AuAu collisions
 at $\theta_{Lab} = 44^\circ$
and \Elab = 1 GeV/u for the parameter-set NL6.
The thin solid, solid, dotted curves indicate the contributions from
the $N + N$, $N + \Delta$  and $\Delta + \Delta$  collisions while
the dashed curve shows the total result.
The asterisks show the experimental data \cite{EXPK}.

\item[Fig. 11]
The number of produced \Kp- mesons versus the nuclear density
(in units of $\rho_0$) (a) and versus the baryon effective mass
( in units of the bare nucleon mass) (b)
in \AuAu collisions at 1 GeV/u for b = 0 fm.
The solid, dashed, dotted and thick dotted lines show the results for POL6,
NL6, NL7 and NL21, respectively.

\eject

\end{description}

\end{document}